\documentclass[journal, ,x11names]{IEEEtran}

\usepackage{setspace}

\usepackage{subfigure}
\usepackage{amsmath,amsfonts}
\usepackage{algorithmic}
\usepackage{algorithm}
\usepackage{array}
\usepackage{subfig}
\usepackage{textcomp}
\usepackage{stfloats}
\usepackage{url}
\usepackage{verbatim}
\usepackage{graphicx}
\usepackage{cite}
\usepackage{subfloat}
\usepackage{tikz}
\usepackage{bm}

\usepackage{color,xcolor}
\usepackage{caption}
\usepackage{pgfplots}
\usetikzlibrary{calc,arrows.meta,positioning}
\usetikzlibrary{shapes.arrows}
\usetikzlibrary{fit,backgrounds} 
\usepackage{standalone}
\usepackage[numbers,sort&compress]{natbib}
\usepackage{setspace}

\begin{document}

\title{Multi-user Wireless Image Semantic Transmission over MIMO Multiple Access Channels}

\author{Bingyan Xie, Yongpeng Wu,~\IEEEmembership{Senior Member,~IEEE,} Feng Shu,~\IEEEmembership{Member,~IEEE,}\\ Jiangzhou Wang,~\IEEEmembership{Fellow,~IEEE,} Wenjun Zhang,~\IEEEmembership{Fellow,~IEEE}

\thanks{(Corresponding author: Yongpeng Wu.)}
\thanks{Bingyan Xie is with the Department of Electronic Engineering, Shanghai Jiao Tong University, Shanghai 200240, China, and also with the 6G R$\&$D Department, ZGC Institute of Ubiquitous-X Innovation and Applications, Beijing 100083, China (e-mail: bingyanxie@sjtu.edu.cn).}
\thanks{Yongpeng Wu, and Wenjun Zhang are with the Department of Electronic Engineering, Shanghai Jiao Tong University, Shanghai 200240, China (e-mail:yongpeng.wu, zhangwenjun@sjtu.edu.cn).}
\thanks{Feng Shu is with the School of Information and Communication Engineering and Collaborative Innovation Center of Information Technology, Hainan University, Haikou 570228, China, and also with the School of Electronic and Optical Engineering, Nanjing University of Science and Technology, Nanjing, 210094, China. (e-mail: shufeng0101@163.com)}
\thanks{Jiangzhou Wang is with the School of Engineering, University of Kent, CT2 7NT Canterbury Kent, U.K. (e-mail: j.z.wang@kent.ac.uk).}
}


\maketitle

\begin{abstract}
This paper focuses on a typical uplink transmission scenario over multiple-input multiple-output multiple access channel (MIMO-MAC) and thus propose a multi-user learnable CSI fusion semantic communication (MU-LCFSC) framework. It incorporates CSI as the side information into both the semantic encoders and decoders to generate a proper feature mask map in order to produce a more robust attention weight distribution. Especially for the decoding end, a cooperative successive interference cancellation procedure is conducted along with a cooperative mask ratio generator, which flexibly controls the mask elements of feature mask maps. Numerical results verify the superiority of proposed MU-LCFSC compared to DeepJSCC-NOMA over 3 dB in terms of PSNR.

\end{abstract}


\section{Introduction}\label{s1}
\IEEEPARstart{N}owadays, various modalities of data have emerged in one’s daily lives. The ever continuously expanding data streams have called for urgent requirements for a new efficient and highly compressible paradigm for the future sixth-generation (6G) communications. Semantic communication, which mainly focuses on the inner semantic meanings of data sources rather than accurate bit recovery, becomes potential for many application scenarios, e.g. automatic driving, unmanned aerial vehicle, and augmented reality. In this way, such intelligent semantic-aware techniques reduce communication overhead to a great extent.

The construction of semantic communication frameworks are mainly based on the joint source-channel coding (JSCC), which utilizes deep learning (DL)-based networks to build the semantic codec for data transmission [1-3]. For example, Xie et al. \cite{DeepSC} proposed a Transformer-based DL-enabled semantic communication (DeepSC) framework for text semantic transmission. Dai et al. \cite{NTSCC} blended nonlinear transform coding into the JSCC to adaptively allocate transmission rate and thus provided a nonlinear source-channel coding framework for image semantic transmission.

However, existing semantic communication works mainly focus on the point-to-point wireless transmission, hindering the applications in the broader multi-user scenarios. There are also works [4-6] concentrated on the multi-user semantic communications. Zhang et al. \cite{IS-SNOMA} considered a semantic-bit coexisting system with multiple users and thus proposed a semantic-aware interference-suppressed technique for users in downlink non-orthogonal multiple access (NOMA) scenarios. Li et al. \cite{NOMASC} proposed a NOMA-enhanced semantic communication framework, namely NOMASC, considering the two-user pair and conducting successive interference cancellation (SIC) at the decoding end. Yilmaz et al. \cite{Deep-NOMA} proposed a distributed DeepJSCC over a multiple access channel (MAC), called DeepJSCC-NOMA. A joint decoder was utilized to decouple both the transmitted symbols of two users. 

With the above multi-user semantic communication frameworks, few works consider the semantic transmission under sophisticated MIMO-MACs. Moreover, how to utilze the feedback MIMO channel state information (CSI) to boost the performance of multi-user communication system has not been solved as well. In this paper, we consider a typical two-user pair in NOMA scenario over MIMO-MACs. Inspired by the CSI fusion-based semantic coding designs in [1], which integrate MIMO CSI as side information into the semantic encoder to produce robust semantic codewords against single-user MIMO channels, we further adopt the CSI fusion method into the cooperative semantic decoding stage. A cooperative mask ratio generator is also proposed to adaptively produce corresponding attention mask maps for alleviating the inter-user interference during the SIC process drived by DL networks. The main contributions are as follows

\begin{enumerate}
\item{\textbf{MU-LCFSC Framework:}}
We propose a multi-user learnable CSI fusion semantic communication (MU-LCFSC) framework to conduct the uplink image transmission with two-user pair over MIMO-MAC. MIMO CSI of each user are treated as side information and embedded both in the semantic encoder and decoder to produce proper attention mask maps, so as to mitigate the performance degradation brought by the MIMO-MAC fading and interference.
\item{\textbf{Cooperative Semantic Decoder:}}
We propose a cooperative semantic decoder at the decoding end to perform the successive interference cancellation for each user. The user with strong allocated power is decoded first. Then the decoded results from stronger user are substracted for the latter successive decoding of weaker user. To adaptively control the mask ratio of attention mask maps for each successive decoder, we further construct the cooperative mask ratio generator. Along with the generated mask ratio range and mask ratio selection vector, the suitable mask ratio can be acquired.

\end{enumerate}

\begin{figure*}[htbp]
	\centering
	\includegraphics[width=6.2in]{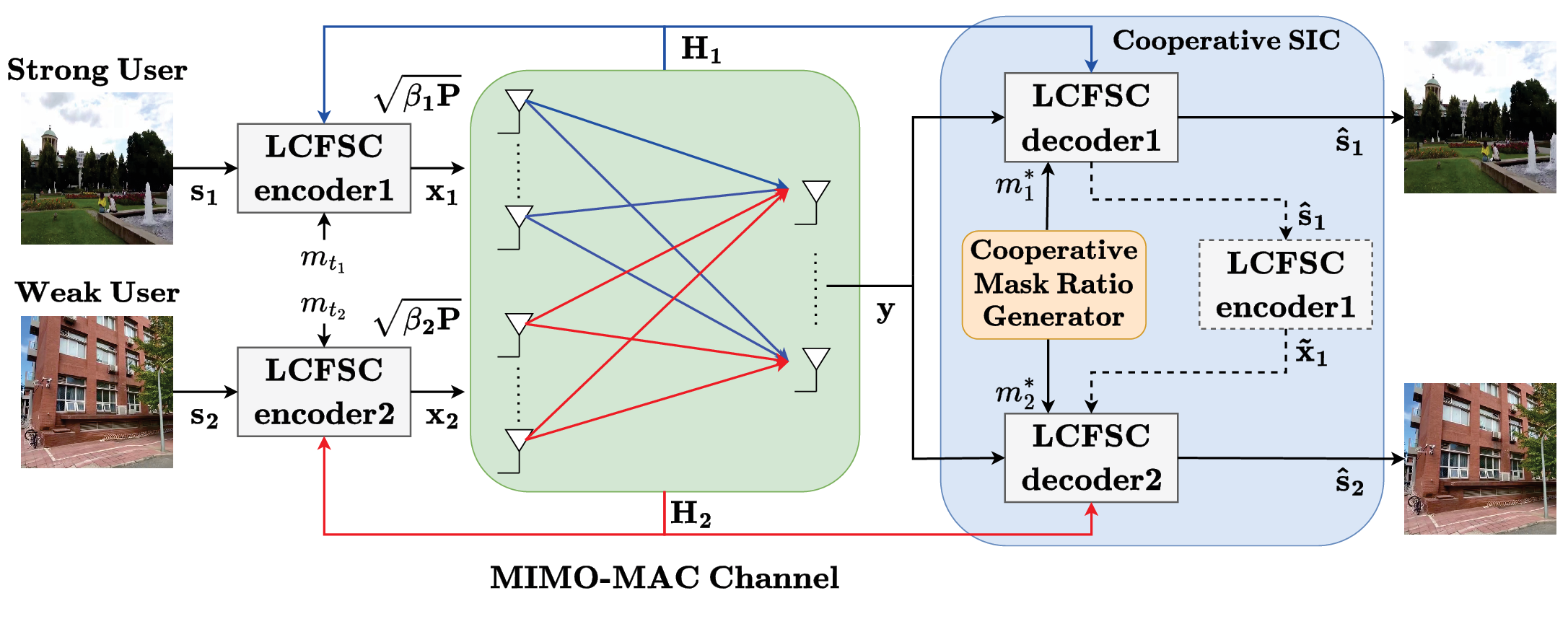}
	\caption{MU-LCFSC framework with two-user pair over MIMO-MAC channels. Two users independently encode their images and decode them successively. The strong user decodes the codewords first and then the weak user utilizes the reconstructed results of strong user to recover the image.}
	\label{fig_1}
\end{figure*}

Notational Conventions: $\mathbb{R}$ and $\mathbb{C}$ refer to the real and complex number sets, respectively. $\mathcal{N}\left (\mu, \sigma^2 \right)$ denotes a Gaussian distribution with mean $\mu$ and variance $\sigma^2$. $\mathbb{E}$ refers to the mathematical expectation. $\odot$ represents element-wise multiplication. Finally, $\left(\cdot\right)^{T}$ denotes the matrix transpose.

\section{System Model and Proposed Framework}

\subsection{System Model}
As shown in Fig. \ref{fig_1}, consider a typical uplink wireless image transmission problem with a two-user pair under MIMO-MAC channels. Given a set with $N$ different images of the $i$-th user $\mathcal{S}_i=\left\{\mathbf{s}_{i,1},\mathbf{s}_{i,2},\cdots,\mathbf{s}_{i,N}\right\}, i = 1,2$, where each image $\mathbf{s}_{i,j}\in\mathbb{R}^{H\times W\times 3}$. Each semantic encoder at the transmitting end encodes the image set $\mathcal{S}_i$ into a codeword sequence set $\mathcal{X}_i=\left\{\mathbf{x}_{i,1},\mathbf{x}_{i,2},\cdots,\mathbf{x}_{i,N}\right\}$, where $\mathbf{x}_{i,j}\in \mathbb{R}^{C_{\mathrm{L}}}$ is the transmitted codewords of the $j$-th image with length $C_{\mathrm{L}}$. In this way, channel bandwidth ratio (CBR) is defined as $R=\frac{C_\mathrm{L}}{H\times W\times 3}$. After that, the codewords pass through the MIMO-MACs, which can be formulated as
\begin{align}
	\mathbf{y}_j=\sqrt{\beta_1P}\mathbf{H}_1\mathbf{x}_{1,j}+\sqrt{\beta_2P}\mathbf{H}_2\mathbf{x}_{2,j}+\mathbf{z}, 
\end{align}
where $P$ is the total transmission power, $\sqrt{\beta_1}$ and $\sqrt{\beta_2}$ are the power allocation factors $(\beta_1+\beta_2=1, \beta_1>\beta_2)$, $\mathbf{x}_{1,j}, \mathbf{x}_{2,j}\in\mathbb{R}^{N_{\mathrm{T}}\times \frac{C_{\mathrm{L}}}{N_{\mathrm{T}}}}$ are the reshaped codewords for MIMO transmission, $\mathbf{y}_j\in\mathbb{R}^{N_{\mathrm{R}}\times \frac{C_{\mathrm{L}}}{N_{\mathrm{T}}}}$ is the received codewords, $\mathbf{H}_1, \mathbf{H}_2\in\mathbb{C}^{N_{\mathrm{R}}\times N_{\mathrm{T}}}$ are the practical MIMO channel state information matrices while $\mathbf{z}\in\mathbb{C}^{N_{\mathrm{R}}\times \frac{C_{\mathrm{L}}}{N_{\mathrm{T}}}}$ is a complex Gaussian noise matrix with mean $0$ and variance $\sigma^2$ of each element.

Finally, the decoder at the base station translates the transmitted codewords into each reconstructed image set $\hat{\mathcal{S}}_i=\left \{\hat{\mathbf{s}}_{i,1},\hat{\mathbf{s}}_{i,2},\cdots,\hat{\mathbf{s}}_{i,N}\right\}$.

\subsection{Proposed Framework of MU-LCFSC}
The proposed MU-LCFSC framework is shown in Fig. \ref{fig_1}. Each semantic extractor, namely LCFSC encoder \cite{LCFSC}, $f_{e_i}(\cdot, \cdot, \cdot): \mathbb{R}^{H\times W\times C}\times\mathbb{C}^{N_{\mathrm{R}}\times N_{\mathrm{T}}}\times{[0,1]} \mapsto \mathbb{R}^{C_{\mathrm{L}}}$, encodes the original images, $\mathbf{s}_{i}$, aided by side information including MIMO CSI, $\mathbf{H}_{i}$, and a hyper-parameter called semantic mask ratio, $m_{t_i}$, into semantic features.

At the decoder end, we conduct the cooperative successive interference cancellation (C-SIC) with the help of a joint cooperative mask ratio generator (CMRG) to flexibly adjust the mask element percentage of attention weight maps. For the stronger user, the reconstructed images, $\mathbf{\hat{s}}_{1}$, are directly generated by the LCFSC decoder, $f_{d_1}(\cdot, \cdot, \cdot): \mathbb{R}^{C_{\mathrm{L}}}\times\mathbb{C}^{N_{\mathrm{R}}\times N_{\mathrm{T}}}\times{[0,1]} \mapsto \mathbb{R}^{H\times W\times C}$. Then with $\mathbf{\hat{s}}_{1}$ from the stronger user, the weaker user substracts the previous decoded images and then decodes its own images $\mathbf{\hat{s}}_{2}$ with $f_{d_2}(\cdot, \cdot, \cdot)$. The received codewords $\mathbf{y}_{s}$ for $f_{d_2}(\cdot, \cdot, \cdot)$ can be formulated as
\begin{align}
	\mathbf{y}_{s} & = \mathbf{y}-\sqrt{\beta_2P}f_{e_1}(\mathbf{\hat{s}}_1,\mathbf{H}_1,m_{t_1})
\end{align}
The whole decoding process is presented below
\begin{align}
\begin{matrix}
	\mathbf{\hat{s}}_1 & = f_{d_1}(\mathbf{y},\mathbf{H}_1,m_{1}^\ast)\\
	\mathbf{\hat{s}}_2 & = f_{d_2}(\mathbf{y}_{s},\mathbf{H}_2,m_{2}^\ast)
\end{matrix}
\end{align}
where $m_{1}^\ast$ and $m_{2}^\ast$ are the semantic mask ratios for LCFSC decoders.

\section{Cooperative Decoding Designs for the Multi-user Detection}

\begin{figure}[htbp]
	\centering
	\includegraphics[width=3.0in]{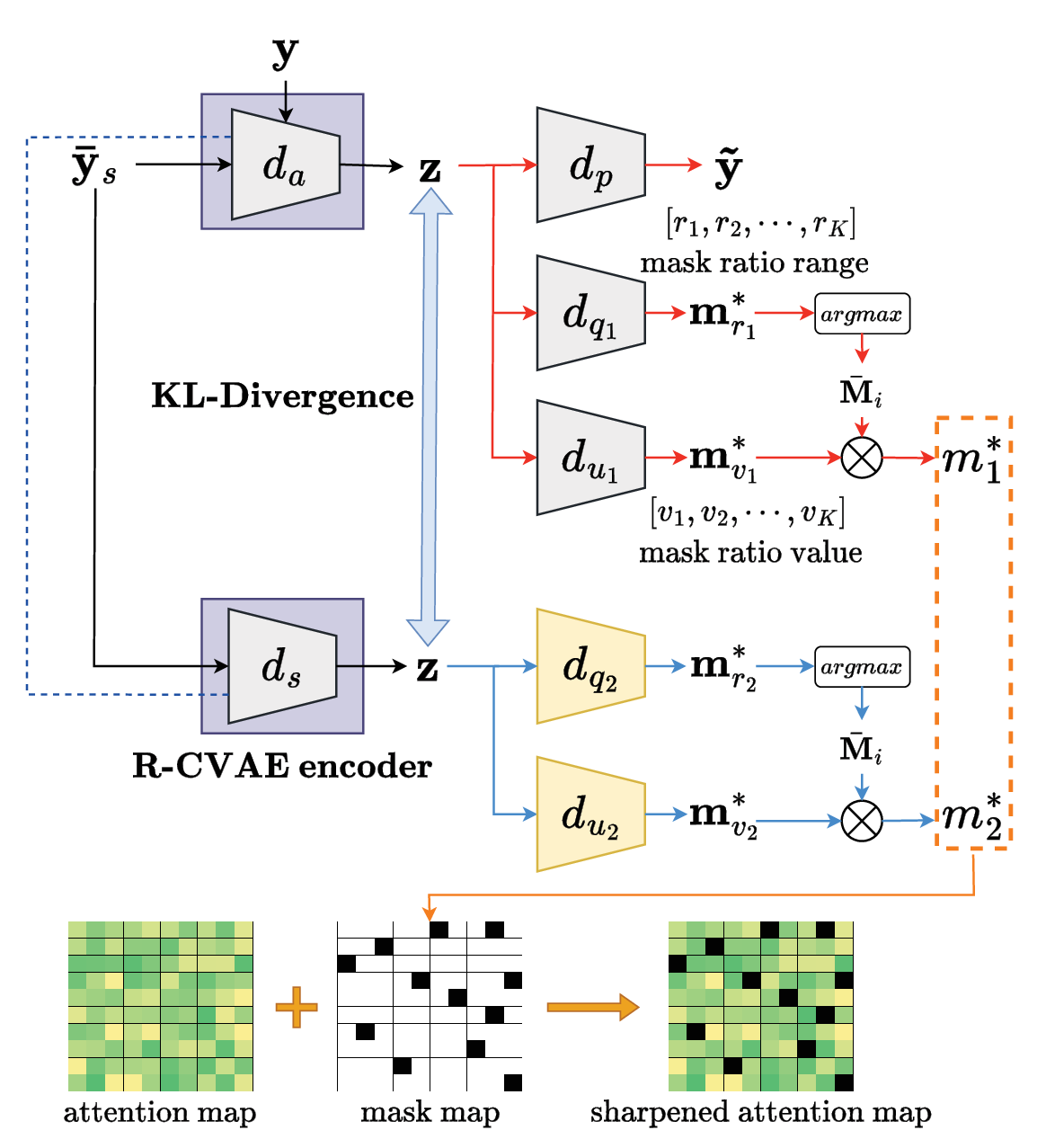}
	\caption{The structure of cooperative mask ratio generator. Each mask ratio is learnt jointly through the chosen mask ratio range and value.}
	\label{fig_2}
\end{figure}
In section \uppercase\expandafter{\romannumeral+2}, the decoder of MU-LCFSC utilizes the C-SIC to decouple the mixed received semantic codewords from two users. An extra CMRG is deployed in the decoder part for providing suitable learnable mask ratio, $\mathbf{m}_{i}^\ast$, to control the mask percentage of elements in attention weight maps. Intuitively, since the stronger user utilizes initial mixed received semantics $\mathbf{y}$ for translating $\mathbf{s}_1$, much severer inter-user interference requires a larger semantic mask ratio to alleviate the performance degradation. While for the weaker user, due to the substraction of the stronger user's signal, less elements in attention weights are obliged to be masked for mitigating interference. Such obvious different channel states between two users would pose difficulty for the mask ratio adaptation, which is hard for the R-CVAE in \cite{LCFSC} to tackle. As such, a unified adaptation for various mask ratios is required.

The structure of CMRG is given in Fig. \ref{fig_2}. Based both on the conditional variational generation \cite{CVAE} and SIC, we treat $\bar{\mathbf{y}}_{s}$ as the auxiliary condition produced the same as $\mathbf{y}_{s}$ by the previous fixed network weight while $\mathbf{y}$ as input observation data to generate proper semantic mask ratios. The conditional log-likelihood can be written as
\begin{equation}
	\begin{aligned}
		\log_{}{p}(\mathbf{y}|\bar{\mathbf{y}}_{s}) & = \mathbb{E}_{\mathbf{z}\sim q(\mathbf{z}|\mathbf{y},\bar{\mathbf{y}}_{s})}[\log_{}{p}(\mathbf{y},\mathbf{z}|\bar{\mathbf{y}}_{s})-\log_{}{p}(\mathbf{z}|\mathbf{y},\bar{\mathbf{y}}_{s})] \\
		& \overset{\text{(a)}}{=} D_{\mathrm{KL}}[q(\mathbf{z}|\mathbf{y},\bar{\mathbf{y}}_{s})||p(\mathbf{z}|\mathbf{y},\bar{\mathbf{y}}_{s})] \\ & + \mathbb{E}_{\mathbf{z}\sim q(\mathbf{z}|\mathbf{y},\bar{\mathbf{y}}_{s})} [\log_{}{p}(\mathbf{y},\mathbf{z}|\bar{\mathbf{y}}_{s})-\log_{}{q}(\mathbf{z}|\mathbf{y},\bar{\mathbf{y}}_{s})]\\  
		& \ge  \mathbb{E}_{\mathbf{z}\sim q(\mathbf{z}|\mathbf{y},\bar{\mathbf{y}}_{s})} [\log_{}{p}(\mathbf{y},\mathbf{z}|\bar{\mathbf{y}}_{s})-\log_{}{q}(\mathbf{z}|\mathbf{y},\bar{\mathbf{y}}_{s})],
	\end{aligned}
\end{equation}
where the first term $D_{\mathrm{KL}}[q(\mathbf{z}|\mathbf{y},\bar{\mathbf{y}}_{s})||p(\mathbf{z}|\mathbf{y},\bar{\mathbf{y}}_{s})]$ in $(\text{a})$ represents the differences between the true posterior and the approximation posterior distribution. The second term is named evidence lower bound (ELBO), which can be rewritten as
\begin{equation}
	\begin{aligned}
		\mathrm{ELBO} = & \mathbb{E}_{\mathbf{z}\sim q(\mathbf{z}|\mathbf{y},\bar{\mathbf{y}}_{s})}[\log_{}{p}(\mathbf{y},\mathbf{z}|\bar{\mathbf{y}}_{s})-\log_{}{q}(\mathbf{z}|\mathbf{y},\bar{\mathbf{y}}_{s})] \\ = & \mathbb{E}_{\mathbf{z}\sim q(\mathbf{z}|\mathbf{y},\bar{\mathbf{y}}_{s})}[\log_{}{p}(\mathbf{y}|\mathbf{z},\bar{\mathbf{y}}_{s})]\\ -& D_{\mathrm{KL}}[q(\mathbf{z}|\mathbf{y},\bar{\mathbf{y}}_{s})||p(\mathbf{z}|\bar{\mathbf{y}}_{s})].
	\end{aligned}
\end{equation}

From the last equation, we observe that the ELBO can be rewritten as the sum of two terms. The first term encapsulates the distortion, when reconstructed from the encoding $\mathbf{z}$ along with condition $\bar{\mathbf{y}}_{s}$. The second one is a regulation term that ensures the latent variables given $\mathbf{y}$ and $\bar{\mathbf{y}}_{s}$ being close to the corresponding encoding given $\bar{\mathbf{y}}_{s}$.

As the reparametrization trick is adopted to produce the latent variables $\mathbf{z}$, we define that conditioned on $\bar{\mathbf{y}}_{s}$, $\mathbf{z}$ is normally distributed with mean $f_{\mu}(\mathbf{y})$ and a diagonal covariance matrix with ${\exp}({f_{\sigma}(\mathbf{y})})$ as diagonal entries. The posterior distribution of $\mathbf{z}$ given $\mathbf{y}$ and $\bar{\mathbf{y}}_{s}$ are approximated by a normal Gaussian distribution with mean $h_{\mu}(\mathbf{y},\bar{\mathbf{y}}_{s})$ and a diagonal covariance matrix with ${\exp}({h_{\sigma}(\mathbf{y},\bar{\mathbf{y}}_{s})})$. In this way, the latent variables $\mathbf{z}$ can be given as
\begin{align}
	\mathbf{z}=h_{\mu}(\mathbf{y},\bar{\mathbf{y}}_{s})+\bm{\epsilon}\odot h_{\sigma}(\mathbf{y},\bar{\mathbf{y}}_{s}),
\end{align}
where $\bm{\epsilon} \sim \mathcal{N}(0,\bm{I})$ denote the sampled normal Gaussian variables.

Along with the CMRG structure, with the acquired latent variables $\mathbf{z}$, the mask ratio range selection vector $\mathbf{m}_{r_i}^{\ast}=[r_1,r_2,\cdots,r_K]\in\mathbb{R}^{K}$ and the mask ratio value selection vector $\mathbf{m}_{v_i}^{\ast}=[v_1,v_2,\cdots,v_K]\in\mathbb{R}^{K}$ are learned simultaneously, in which each value represents the weight of range selection and value selection, respectively. The predefined mask ratio range is denoted as $\bar{\mathbf{M}}_i=[m_1,m_2,\cdots,m_K]\in\mathbb{R}^{K}, i=1,\cdots,K$. The final semantic mask ratio $m_i^\ast$ can be computed as 
\begin{align}
	m_i^\ast & = \bar{\mathbf{M}}_{argmax(\mathbf{m}_{r_i}^{\ast})}{\mathbf{m}_{v_i}^{\ast}}^T.
\end{align}
where $argmax(\cdot)$ denotes the serial number of the maximum element in the vector.

\section{Implementation Details}
Based on the above analysis, we present the training loss function for the MU-LCFSC.

For wireless image transmission, we denote $L_\mathrm{1}$ as the image reconstruction loss for both users, which can be expressed as
\begin{align}
	L_\mathrm{1}=\frac{1}{2N}\sum_{i=1}^{2}\sum_{j=1}^{N}||\mathbf{\hat{s}}_{i,j} - \mathbf{s}_{i,j}||^2,
\end{align}
where $2N$ refers to the total number of source images, $||\cdot||^2$ is the mean square error (MSE) loss function.

For the learnable mask ratio generation, the encoder part is the same as LCFSC framework, expressed as
\begin{align}
	L_{\mathrm{c}} = L_{\mathrm{c}_1} + L_{\mathrm{c}_2}. 
\end{align}
where $L_{\mathrm{c}_1}$ and $L_{\mathrm{c}_2}$ represent the corresponding condition generation loss in \cite{LCFSC} of each user, respectively.

For the decoder part, with the CMRG, the reconstruction loss and recongition loss can be written as
\begin{align}
	L_{\mathrm{rec}} = \mathbb{E}_{\mathbf{z}\sim q(\mathbf{z}|\mathbf{y},\bar{\mathbf{y}}_{s})}[\log_{}{p}(\mathbf{y}|\mathbf{z},\bar{\mathbf{y}}_{s})]=\frac{1}{N}\sum_{j=1}^{N}||\tilde{\mathbf{y}}_j-{\mathbf{y}}_j||^2,
\end{align}
\begin{equation}
	\begin{aligned}
		L_{\mathrm{reg}} = & D_{\mathrm{KL}}[q(\mathbf{z}|\mathbf{y},\bar{\mathbf{y}}_{s})||p(\mathbf{z}|\bar{\mathbf{y}}_{s})] \\ = &  \frac{1}{N}\sum_{j=1}^{N}\sum_{i=1}^{L}\bigg[f_{ji\sigma}(\mathbf{y})-h_{ji\sigma}(\mathbf{y},\bar{\mathbf{y}}_{s}) \\+ & \exp(h_{ji\sigma}(\mathbf{y},\bar{\mathbf{y}}_{s})-f_{ji\sigma}(\mathbf{y}))\\+& \left.\frac{[f_{ji\mu}(\mathbf{y})-h_{ji\mu}(\mathbf{y},\bar{\mathbf{y}}_{s})]^2}{\exp(f_{ji\sigma(\mathbf{y})})} \right], 
	\end{aligned}
\end{equation}
where $\tilde{\mathbf{y}}_{j}$ represents the reconstructed codewords of the $j$-th image, $f_{ji\mu}$ and $h_{ji\mu}$ denote the $i$-th mean element of the function $f$ and $h$ while $f_{ji\sigma}$ and $h_{ji\sigma}$ are the $i$-th convariance matrix element of the function $f$ and $h$, respectively. The sequence length of the latent representation is denoted as $L$.

The total loss for CMRG at the decoder is denoted as
\begin{align}
	L_{\mathrm{sic}} = L_{\mathrm{rec}} + L_{\mathrm{reg}}. 
\end{align}

Overall, combining the JSCC and CMRG part together, the training loss of LCFSC is formulated as
\begin{align}
	L_{\mathrm{2}} = L_{\mathrm{1}} + \lambda (L_{\mathrm{c}}+L_{\mathrm{sic}}) 
\end{align}
where $\lambda$ is the trade-off term controlling $L_1$, $L_{\mathrm{c}}$, $L_{\mathrm{sic}}$.

\section{Numerical Results}
In this section, numerical results are presented to verify the effectiveness of MU-LCFSC. 

\subsection{Experimental Setups}

\subsubsection{Datasets}

For the wireless semantic image transmission, we quantify the performances of MU-LCFSC versus other benchmarks over the UDIS-D \cite{UDIS-D} dataset. During model training, images are resized into the shape of 128$\times$128$\times$3.

\subsubsection{Model Deployment Details}
The network deployment of MU-LCFSC is the same according to \cite{LCFSC} based on the Swin-Transformer \cite{Swin} backbone. We set MIMO antenna numbers as $N_\mathrm{T} = 2$ and $N_\mathrm{R} = 2$. Through trial and error, power allocation factors are set as $(\beta_1,\beta_2)=(0.7,0.3)$ and loss trade-off term $\lambda$ as $0.3$. For the uplink transmission of each user, MIMO CSI matrices are generated according to \cite{channel} with 1000 samples of MIMO CSI matrices for training and 100 extra samples for testing, respectively.

\subsubsection{Comparison Benchmarks}
In the experiments, several benchmarks are given as below

$\textbf{WITT}$: Wireless Image Transmission Transformer in \cite{WITT}.

$\textbf{DeepJSCC-NOMA}$: Distributed Deep Joint Source-Channel Coding in \cite{Deep-NOMA} with a single decoder.

$\textbf{LCFSC}$: LCFSC in \cite{LCFSC} where only the encoders adopt the CSI-fusion masking strategy.

$\textbf{MU-LCFSC (OMA)}$: MU-LCFSC transmits images with two independent links of equal power allocation.

\subsubsection{Evaluation Metrics}

We leverage the widely used pixel-wise metric peak signal-to-noise ratio (PSNR) and the perceptual-level multi-scale structural similarity (MS-SSIM) along with learned perceptual image patch similarity (LPIPS) as measurements for the reconstructed image quality.

\subsection{Results Analysis}

\subsubsection{SNR Performances}
\begin{figure*}[htbp]
	\centering  
	\subfigure[PNSR for the reconstructed images.]{
		\includegraphics[width=0.32\linewidth]{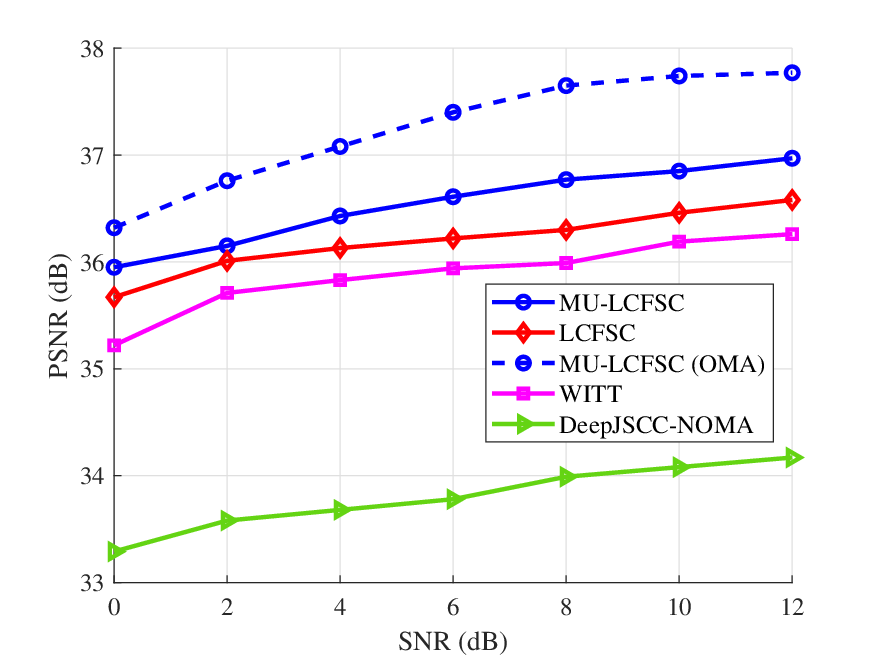}}
	\subfigure[MS-SSIM for the reconstructed images.]{
		\includegraphics[width=0.32\linewidth]{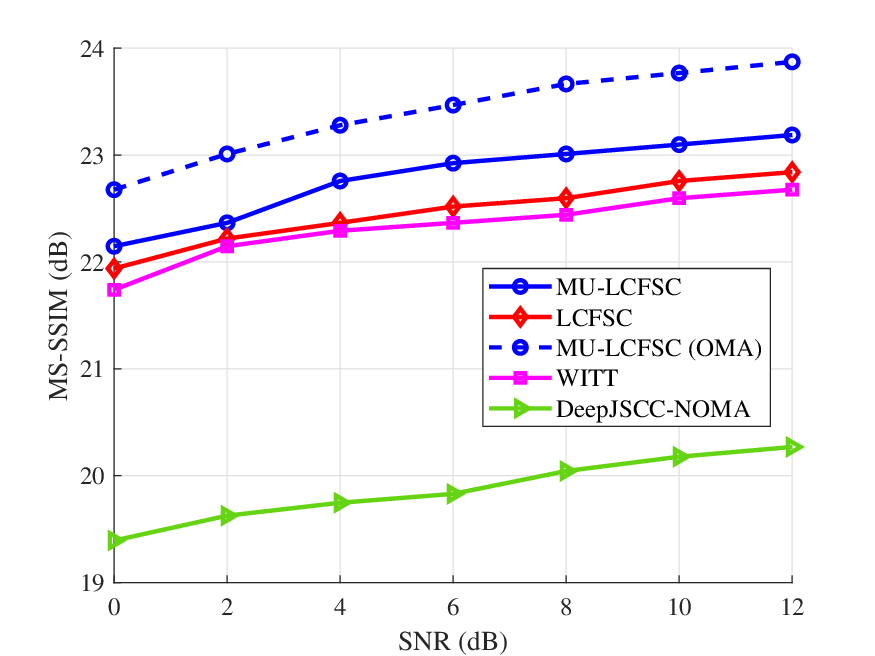}}
	\subfigure[LPIPS for the reconstructed images.]{
		\includegraphics[width=0.32\linewidth]{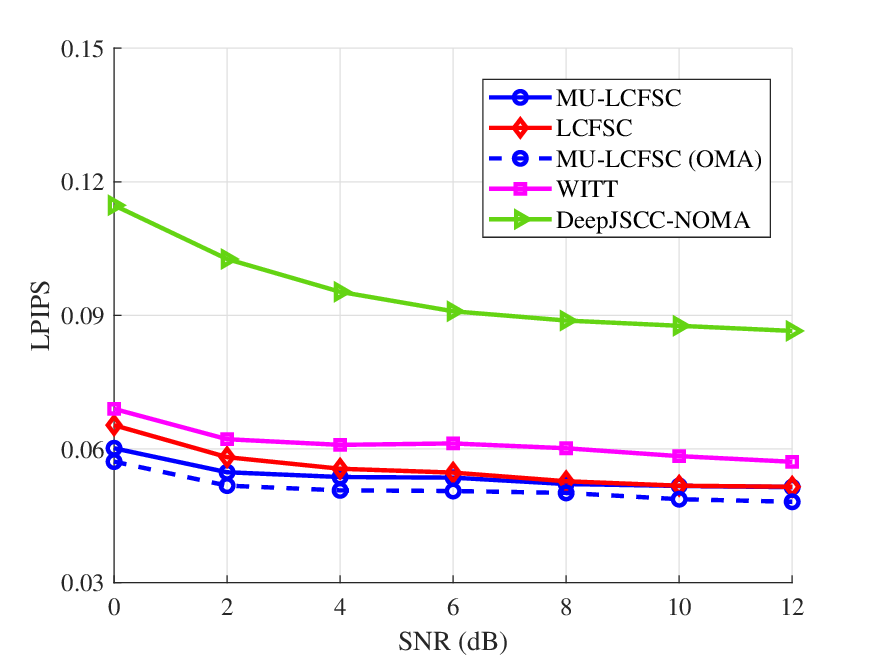}}
	\caption{Quality of the reconstructed images versus the SNRs in MIMO fading channels ($R$ = 0.06).}
	\label{fig_3}
\end{figure*}

We first present the SNR performances for the MU-LCFSC and other benchmarks in Fig. \ref{fig_3}. From Fig. \ref{fig_3}(a), It is seen that MU-LCFSC outperforms WITT in all ranges of SNRs, demonstrating the CSI-aware ability through incorporating CSI as side information into semantic encoders and decoders. For the LCFSC, it can also serve as an ablation study for the proposed MU-LCFSC. The DeepJSCC-NOMA, which utilizes a unified decoder to decouple the reconstructed images of two users at the same time, shows a evident performance gap compared to other schemes which adapt such SIC techniques, generally about 3 dB lower than MU-LCFSC. It is seen that the superposed semantics along with fading and noise can not be easily recovered with a single decoder. Finally, for the MU-LCFSC (OMA), we employ independent links for each user while the transmitting CBR of each user is the same as the NOMA transmission conditions. In this way, such orthogonal transmission scheme pretends to be an upper bound for the MU-LCFSC. The performance gap is limited in about 0.8 dB, which is reasonably satisfying compared to the saving of band resources of NOMA transmission. From Fig. \ref{fig_3}(b) and Fig. \ref{fig_3}(c), the MS-SSIM and LPIPS performance stay the similar trend as the PSNRs. With the adaptive sampled SNRs during training stage, the total performances are satisfying.

\subsubsection{CBR Performances}
\begin{figure*}[htbp]
	\centering  
	\subfigure[PNSR for the reconstructed images.]{
		\includegraphics[width=0.32\linewidth]{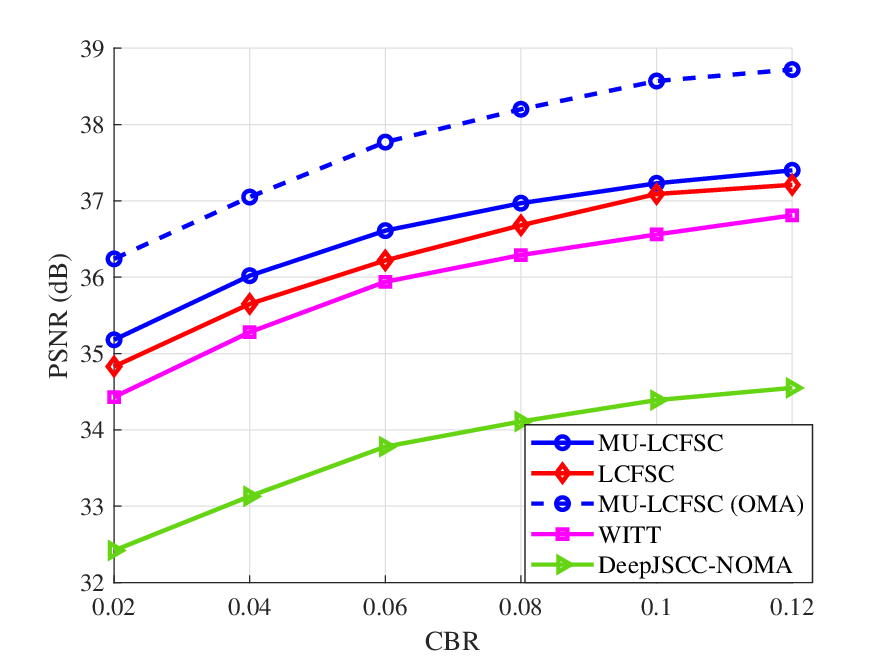}}
	\subfigure[MS-SSIM for the reconstructed images.]{
		\includegraphics[width=0.32\linewidth]{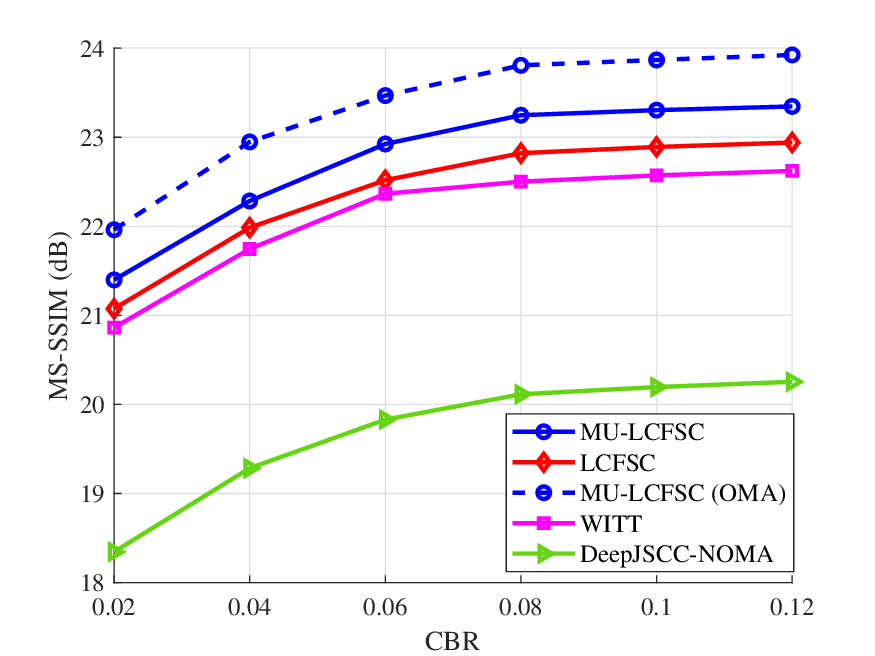}}
	\subfigure[LPIPS for the reconstructed images.]{
		\includegraphics[width=0.32\linewidth]{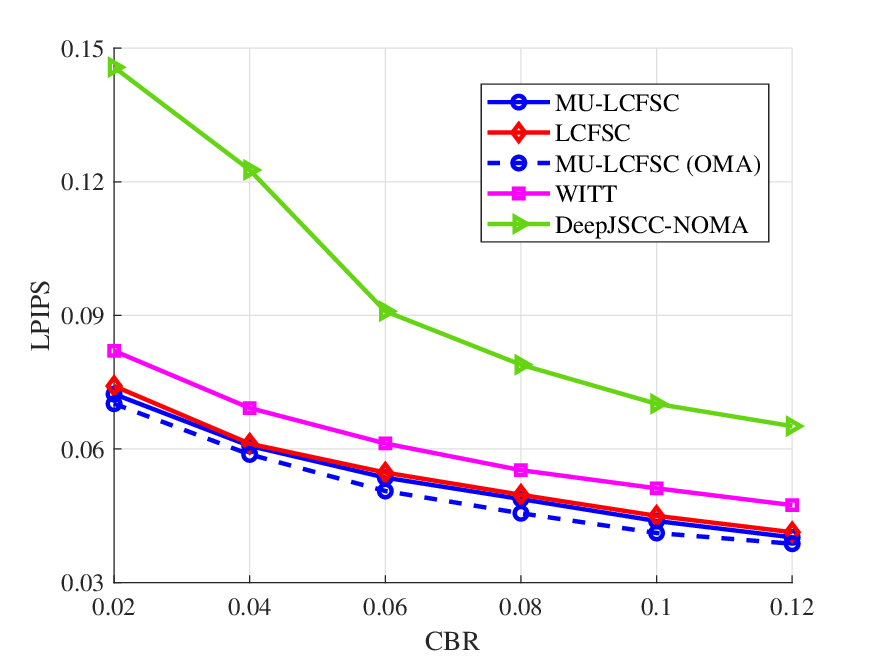}}
	\caption{Quality of the reconstructed images versus the CBRs in MIMO fading channels (SNR = 6 dB).}
	\label{fig_4}
\end{figure*}

Then we evaluate the CBR performances in Fig. \ref{fig_4}. In summary, the MU-LCFSC generally outperforms other DL-based schemes in all CBRs for both PSNR and MS-SSIM metrics in NOMA transmission scenarios. Even in extreme low CBR such as 0.02 or 0.04, LCFSC still achieves relatively satisfying performances, which indicates the superiority of utilizing CSI-aware codec structure for the efficient data compression and transmission. Since MU-LCFSC enables adaptively adjusting the source and channel coding rate based on deep JSCC structure while ensuring the CSI-aware performances through robust semantic coding and cooperative SIC decoding, it performs to be efficient in different channel bandwidth conditions.

\subsubsection{Visualization Results for the Wireless Video Transmission}
\begin{figure}[htbp]
	\centering
	\includegraphics[width=2.9in]{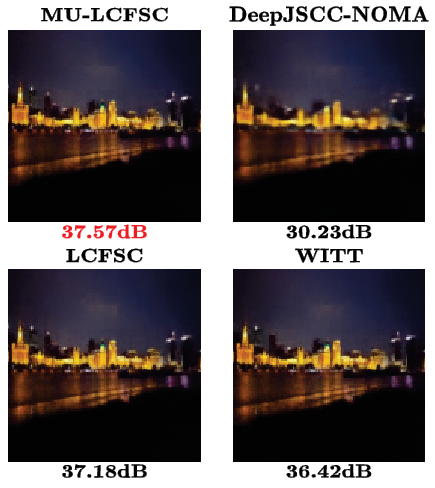}
	\caption{Visualized results for the MU-LCFSC and other benchmarks. (SNR = 12 dB, CBR = 0.10)}
	\label{fig_5}
\end{figure}

Finally, we present the visualization results in Fig. \ref{fig_5}. For other DL-based schemes such as LCFSC and WITT, MU-LCFSC achieves better PSNR performances. For the DeepJSCC-NOMA, obvious blurry areas exist, which illustrates the drawback of such decoding structure with a single decoder. With proposed MU-LCFSC, reconstructed images with sound visual reconstructed quality are provided.

\subsubsection{Complexity Analysis}
Finally, we analyse the complexity of proposed MU-LCFSC. As shown in Tab. \ref{table1}, with extra proposed CMRG part, MU-LCFSC has higher Parameters but competitive computation cost compared to LCFSC. If the number of users in NOMA scenarios increases, the performance gap between MU-LCFSC and LCFSC would be enlarged. It turns to be a trade-off between model parameters and transmission accuracy.

\begin{table}[htbp]
	\centering
	\caption{Evaluation of complexity and computation cost.}
	\label{table1}
	
	\begin{tabular}{|c|c|c|c|}  
		\hline 
		& &\\[-6pt] 
		Metric&FLOPs (G)&Parameters (M) \\
		\hline
		& &\\[-6pt]  
		MU-LCFSC&47.7&511.2 \\
		\hline
		& &\\[-6pt]  
		LCFSC&46.5&287.8 \\
		\hline
	\end{tabular}
\end{table}

\fontsize{8pt}{10pt}\selectfont

\begin{thebibliography}{50}

\bibitem{DeepSC}
H.~Xie, Z.~Qin, G.~Y. Li, and B.-H. Juang, ``Deep learning enabled semantic communication systems'', \emph{IEEE Trans. Signal Process.}, vol.~69, pp. 2663--2675, Apr. 2021.

\bibitem{NTSCC}
J. Dai et al., ``Nonlinear Transform Source-Channel Coding for Semantic Communications", \emph{IEEE J. Select. Areas Commun.}, vol. 40, no. 8, pp. 2300-2316, Aug. 2022.

\bibitem{DeepWiVe}
T. -Y. Tung and D. Gündüz, "DeepWiVe: Deep-Learning-Aided Wireless Video Transmission," in \emph{IEEE J. Select. Areas Commun.}, vol. 40, no. 9, pp. 2570-2583, Sept. 2022.

\bibitem{IS-SNOMA}
Y. Zhang, R. Zhong, Y. Liu, W. Xu and P. Zhang, "Interference Suppressed NOMA for Semantic-aware Communication Networks," \emph{IEEE Trans. Wire. Commun.}, (early access), Mar., 2024.

\bibitem{NOMASC}
W. Li, H. Liang, C. Dong, X. Xu, P. Zhang and K. Liu, "Non-Orthogonal Multiple Access Enhanced Multi-User Semantic Communication," \emph{IEEE Trans. Cognit. Commun. Networking}, vol. 9, no. 6, pp. 1438-1453, Dec. 2023.

\bibitem{Deep-NOMA}
S. F. Yilmaz, C. Karamanlı and D. Gündüz, "Distributed Deep Joint Source-Channel Coding over a Multiple Access Channel," \emph{IEEE Int. Conf. on Commun. (ICC)}, Rome, Italy, May 2023, pp. 1400-1405.

\bibitem{LCFSC}
B.~Xie, Y.~Wu, Y.~Shi, W.~Z, S.~Cui and M.~Debbah, "Robust Image Semantic Coding with Learnable CSI Fusion Masking over MIMO Fading Channels," \emph{arxiv:2406.07389}, May 2024. [Online]. Available: \url{https://arxiv.org/abs/2406.07389}.

\bibitem{CVAE}
G.~Pandey, A.~Dukkipati, ``Variational methods for conditional multimodal deep learning", \emph{Int. Jt. Conf. Neural Networks (IJCNN)}, Anchorage, AK, USA, 2017, pp. 308-315.

\bibitem{UDIS-D}
L.~Nie, C.~Lin, K.~Liao, et al. ``Unsupervised deep image stitching: Reconstructing stitched features to images'', \emph{IEEE Trans.	Image Process.}, vol. 30, pp. 6184--6197, Jul. 2021.

\bibitem{Swin}
Z.~Liu, Y.~Lin, Y.~Cao, et al. ``Swin transformer: Hierarchical vision transformer using shifted windows", \emph{IEEE Conf. Comput. Vis. Pattern Recognit. (CVPR)}, Montreal, QC, Canada, Oct. 2021, pp. 9992-10002.

\bibitem{channel}
S.~Wu, C.~Wang, M.~Alwakeel, et al. ``A general 3-D non-stationary 5G wireless channel model", \emph{IEEE Trans. Commun.}, vol. 66, no. 7, pp. 3065-3078, Jul. 2018.

\bibitem{WITT}
K.~Yang, S.~Wang, J.~Dai, et al. ``WITT: A wireless image transmission transformer for semantic communications", \emph{IEEE Int. Conf. Acoust. Speech Signal Process. (ICASSP)}, Rhodes Island, Greece, Jun. 2023, pp. 1-5.


\end{thebibliography}

\vfill

\end{document}